\documentclass[aps,print,showpacs,twocolumn]{revtex4}%
\usepackage{amsfonts}
\usepackage{amsmath}
\usepackage{amssymb}
\usepackage{graphicx}%
\begin{document}
\author{Gang Chen$^{a,b}$ Juqi Li$^{b}$ J.-Q. Liang$^{b}$}
\affiliation{$^{a}$Department of Physics, Shaoxing College of Arts and Sciences, Shaoxing
312000, China}
\affiliation{$^{b}$Institute of Theoretical Physics, Shanxi University, Taiyuan, Shanxi
030006, China}

\pacs{03.65.Vf, 42.50.Fx}
\title{{\LARGE Critical Property of Geometric Phase in the Dicke Model}}

\begin{abstract}
We obtain the ground-state energy level and associated geometric phase in the
Dicke model analytically by means of the Holstein-Primakoff transformation and
the boson expansion approach in the thermodynamic limit. The non-adiabatic
geometric phase induced by the photon field is derived with the time-dependent
unitary transformation. It is shown that the quantum phase transition
characterized by the non-analyticity of the geometric phase is remarkably of
the first-order. We also investigate the scaling behavior of the geometric
phase at the critical point, which can be measured in a practical experiment
to detect the quantum phase transition.

\end{abstract}
\maketitle

Berry phase\cite{1} with the relaxation of its original restriction
conditions\cite{2,3,4,5} has been extensively generalized along various
directions \cite{6,7,8,9}. Recently, the geometric phase (GP) has been
regarded as an essential way to implement operation of a universal quantum
logic gate in quantum computing\cite{10,11,12,13,14,15} and as an important
tool to detect quantum phase transition (QPT) in the $XY$ model\cite{16,17},
which describes a structural change in the properties of the ground state
energy spectrum associated with the variation of a controlling
parameter\cite{18} and originates from singularity of the energy
spectrum\cite{19}. A single $1/2$-spin driven by a classical rotating magnetic
field is a well studied model to generate the Berry phase. Recently a
quantized magnetic field has been considered and shown to induce the GP which
reduces to the standard Berry phase in the semiclassical limit\cite{20,21}.
The generation of this framework of Berry phase to many-body two-level-atom
system interacting with a single bosonic mode known as Dicke model\cite{22,23}
is certainly of interest particularly in connection with the QPT. The Dicke
model displays an interesting super-radiant phenomenon describing the
collective and coherent behaviors of atoms and exhibits the QPT from the
normal to the super-radiant phases\cite{24,25,26} induced by the variation of
the coupling strength between the atom and field. It has been known that the
atomic ensemble in the normal phase is collectively unexcited while is
macroscopically excited with coherent radiations in the so-called
super-radiant phase. The QPT has been related to the emergence of chaos in a
corresponding classical Hamiltonian\cite{27,28} and the logarithmic diverges
of the von Neumann entropy at the critical transition point which describes
the quantum entanglement between the atoms and field\cite{29,30,31}.

In this paper we derive the non-adiabatic GP of many-body atom-system driven
by the quantized field for the first time. The ground-state and corresponding
GP are obtained with the Holstein-Primakoff transformation\cite{32} and the
boson expansion approach\cite{33} in the thermodynamic limit. We demonstrate
that the GP can serve as a critical criteria to characterize the QPT. The
scaling behavior of the GP at the critical point is also studied.

The Dicke model of $N$ two-level atoms in a single-mode light field is given
in the rotating-wave approximation by%

\begin{equation}
H=\omega a^{+}a+\sum_{i=1}^{N}[\frac{\omega_{0}}{2}\sigma_{z}^{i}%
+\frac{\lambda}{\sqrt{N}}(\sigma_{+}^{i}a+\sigma_{-}^{i}a^{+})]
\end{equation}
where $a$, $a^{+}$ are the photon annihilation and creation operators;
$\sigma_{+}^{j}$ and $\sigma_{-}^{j}$ are the pseudo-spin operators for the
$j-$th atom defined as $\sigma_{\pm}^{j}=\sigma_{x}^{j}\pm i\sigma_{y}^{j}$
with $\sigma_{x}\ $and $\sigma_{y}$ being the Pauli matrices. $\lambda$
denotes the coupling strength between the atom and field and $\omega_{0}$ is
the energy difference between two levels of the atom in the unit $\hbar=1$.
The prefactor $1/\sqrt{N}$ is inserted to have a finite free energy per atom
in the thermodynamical limit $N\rightarrow\infty$. The Hamiltonian eq.(1) is
actually considered in a rotating frame along with the light field and becomes%
\begin{equation}
H^{\prime}(t)=R(t)HR^{+}(t)-iR(t)\frac{dR^{+}(t)}{dt},
\end{equation}
in the laboratory frame with the time-dependent unitary transformation%
\begin{equation}
R(t)=\exp[-i\varphi(t)a^{+}a],
\end{equation}
where $\varphi(t)=\omega t$ denotes the rotating angle. The non-adiabatic GP
induced by the light field can be found by solving the time-dependent
Schr\"{o}dinger equation%
\begin{equation}
i\frac{d\left\vert \Psi_{n}^{\prime}(t)\right\rangle }{dt}=H^{\prime
}(t)\left\vert \Psi_{n}^{\prime}(t)\right\rangle ,
\end{equation}
and the result is%
\begin{align}
\gamma_{n}  &  =i\int_{0}^{2\pi}\left\langle \Psi_{n}^{\prime}(\varphi
(t))\right\vert \frac{d}{d\varphi}\left\vert \Psi_{n}^{\prime}(\varphi
(t))\right\rangle d\varphi\nonumber\\
&  =2\pi\left\langle \Psi_{n}\right\vert a^{+}a\left\vert \Psi_{n}%
\right\rangle .
\end{align}
where $\left\vert \Psi_{n}^{\prime}(\varphi(t))\right\rangle =R(t)|\Psi_{n}>$
with $|\Psi_{n}>$ being the eigenstate of the Hamiltonian $H$ such that
$H|\Psi_{n}>=E_{n}|\Psi_{n}>$, which is equal to the Berry phase formula in
Refs.\cite{20,21}. We are interested in the critical property of the GP
associated with the ground state $|\Psi_{0}>$ of the many-body system. Using
collective giant-spin operators $J_{z}=\sum_{i=1}^{N}\sigma_{z}^{i}$, $J_{\pm
}=\sum_{i=1}^{N}\sigma_{\pm}^{i}$, which satisfy the usual SU(2) commutation
relations $[J_{z},J_{\pm}]=\pm J_{\pm}$ and $[J_{+},J_{-}]=2J_{z}$ with total
spin quantum number $j=N/2$, the Hamiltonian eq.(1) becomes\
\begin{equation}
H=\omega a^{+}a+\frac{\omega_{0}}{2}J_{z}+\frac{\lambda}{\sqrt{N}}%
(J_{+}a+J_{-}a^{+}). \label{6}%
\end{equation}
The system undergoes a QPT at a critical value of the atom-field coupling
strength $\lambda_{c}$ (to be determined below) from the normal phase when
$\lambda<\lambda_{c}$ to the super-radiant phase when $\lambda>\lambda_{c}$ in
the thermodynamical limit. By means of the boson expansion approach and the
Holstein-Primakoff transformation of the collective angular momentum operators
defined as $J_{+}=b^{+}\sqrt{N-b^{+}b},J_{-}=\sqrt{N-b^{+}b}b$ and
$J_{z}=(b^{+}b-N/2)$, where the new boson operators satisfy the commutation
relation $[b,b^{+}]=1$\cite{32}. With this transformation the Hamiltonian becomes%

\begin{align}
H  &  =\omega a^{+}a+\frac{\omega_{0}}{2}(b^{+}b-N/2)+\\
&  \frac{\lambda}{\sqrt{N}}(b^{+}\sqrt{N-b^{+}b}a+\sqrt{N-b^{+}b}%
ba^{+})\nonumber
\end{align}
Following the procedure of Ref.[28], we introduce shifting boson operators
$\ c^{+}\ $and $d^{+}\ $with properly scaled auxiliary parameters $\alpha$ and
$\beta$ such that%
\begin{equation}
c^{+}=a^{+}+\sqrt{N}\alpha\text{, }d^{+}=b^{+}-\sqrt{N}\beta\label{8}%
\end{equation}
to evaluate the critical transition point $\lambda_{c}$ and the ground state
energies of both two phases. It should be noticed that the displacements,
$c^{+}=a^{+}-\sqrt{N}\alpha$ and $d^{+}=b^{+}+\sqrt{N}\beta$ , also lead to
the same result except for the change of sign. Expanding Hamiltonian eq.(7)
with the displacement boson operators $c^{+}$ and $d^{+}$ as power series of
$1/\sqrt{N}$ we can obtain\textbf{ }%
\begin{equation}
H=NH_{0}+N^{1/2}H_{1}+N^{0}H_{2}+\cdot\cdot\cdot
\end{equation}
with%

\[
H_{0}=\omega\alpha^{2}+\frac{\omega_{0}}{2}(\beta^{2}-\frac{1}{2}%
)-2\lambda\alpha\beta\sqrt{k},
\]

\begin{align*}
H_{1}  &  =(-\omega\alpha+\lambda\beta\sqrt{k})(c^{+}+c)+\\
&  \lbrack\frac{\omega_{0}\beta}{2}-\lambda\alpha\sqrt{k}(1-2\beta^{2}%
)](d^{+}+d),
\end{align*}

\begin{align*}
H_{2}  &  =\omega c^{+}c+\frac{\omega_{0}}{2}d^{+}d+\lambda\lbrack\sqrt
{k}d^{+}-\frac{\beta^{2}}{2\sqrt{k}}(d^{+}+d)]c\\
&  +\lambda\lbrack\sqrt{k}d-\frac{\beta^{2}}{2\sqrt{k}}(d^{+}+d)]c^{+}+\\
&  \frac{\lambda\alpha\beta}{2\sqrt{k}}[4d^{+}d+(d^{+})^{2}+d^{2}+\frac
{\beta^{2}}{2k}(d^{+}+d)^{2}],
\end{align*}
where $k=1-\beta^{2}$. The first term of $H$ gives rise to the
Hartree-Bogoliubov ground state energy\cite{34}%

\begin{align}
E_{0} &  =\left\langle \Psi_{0}\right\vert NH_{0}\left\vert \Psi
_{0}\right\rangle \nonumber\\
&  =\left\{
\begin{array}
[c]{l}%
-N\omega_{0}/4\text{ ,\ \ \ \ \ \ \ \ \ \ \ \ \ \ \ \ \ \ \ \ \ \ \ \ }%
\lambda<\lambda_{c}\\
-N[\frac{\lambda^{2}}{4\omega}(1-\delta^{2})+\frac{\omega_{0}\delta}%
{4}]\ ,\text{ }\lambda>\lambda_{c}%
\end{array}
\right.  ,
\end{align}
and the critical value%
\begin{equation}
\lambda_{c}=\sqrt{\omega_{0}\omega/2}%
\end{equation}
where $\left\vert \Psi_{0}\right\rangle =\left\vert 0\right\rangle
_{at}\left\vert 0\right\rangle _{ph}$ is the vacuum of the quasi-boson
operators such that $c\left\vert 0\right\rangle _{ph}=0$ and $d\left\vert
0\right\rangle _{at}=0$, (correspondingly $a\left\vert 0\right\rangle
_{ph}=\sqrt{N}\alpha\left\vert 0\right\rangle _{ph}$, $b\left\vert
0\right\rangle _{at}=-\sqrt{N}\beta\left\vert 0\right\rangle _{at}$) with
$\delta=\omega\omega_{0}/2\lambda^{2}$. The auxiliary parameters%

\begin{equation}
\alpha=\left\{
\begin{array}
[c]{l}%
0\text{ \ \ \ \ \ \ \ \ \ \ , }\lambda<\lambda_{c}\\
\lambda\sqrt{1-\delta^{2}}/2\omega\text{, \ }\lambda>\lambda_{c}%
\end{array}
\right.
\end{equation}

\begin{equation}
\beta=\left\{
\begin{array}
[c]{l}%
0\text{ \ \ \ \ \ \ \ \ \ , }\lambda<\lambda_{c}\\
\sqrt{(1-\delta)/2}\text{, }\lambda>\lambda_{c}%
\end{array}
\right.
\end{equation}
are determined from minimizing the ground state energy eq.(10). In the normal
phase ( $\lambda<\lambda_{c}$ ) the system is essentially in the lower energy
state and is only microscopically excited, whereas above the transition point
both the field and the atomic ensemble acquire macroscopic excitations.

The GP of the many-body system associated with the ground state $\left\vert
\Psi_{0}\right\rangle $ is obtained in terms of the quasi-boson operator
eq.(8) as%

\begin{align}
\gamma_{0}  &  =2\pi\left\langle \Psi_{0}\right\vert a^{+}a\left\vert \Psi
_{0}\right\rangle \\
&  =\left\{
\begin{array}
[c]{l}%
0\text{ \ \ \ \ \ \ \ \ \ \ , }\lambda<\lambda_{c}\\
\frac{\pi N}{2\omega^{2}}(\lambda^{2}-\frac{\lambda_{c}^{4}}{\lambda^{2}%
})\text{ ,\ }\lambda>\lambda_{c}%
\end{array}
\right.  .\nonumber
\end{align}
The scaled GP $\gamma_{0}/N$ and its first-order derivative with respect to
the coupling parameter $\lambda$ is shown in Fig.1 with the resonant condition
$\omega=\omega_{0}=1$. It can be seen that the GP vanishes when $\lambda
<\lambda_{c}$ and increases abruptly with $\lambda$ when $\lambda>\lambda_{c}$
indicating a first-order phase transition at the critical point $\lambda_{c}$.
It is apparent that GP can be used to detect the quantum criticality in
systems described by the Dicke model. The quantum criticality is also shown to
relate the quantum entanglement governed by the von Neumann entropy which has
a cups-like behavior at the critical transition point\cite{29,30}. While the
GP behaves as a step function at the critical point.

The scaling behavior of GP at the critical point can be found in the
thermodynamical limit as%

\begin{equation}
\frac{\gamma_{0}}{N}(\lambda\rightarrow\lambda_{c})=\frac{2\pi\lambda_{c}%
}{\omega^{2}}\left\vert \lambda-\lambda_{c}\right\vert .
\end{equation}
On the other hand, the first-order derivative of the GP diverges linearly with
the atom number $N$ at the transition point $\lambda_{c}$ as%

\begin{equation}
\lim_{N\rightarrow\infty}\frac{d\gamma_{0}}{d\lambda}\mid_{\lambda=\lambda
_{c}}=\frac{2\pi\lambda_{c}}{\omega^{2}}N,
\end{equation}
which is different from the logarithmic divergence of the first-order
derivative of the Berry phase in $XY$ model\cite{16,17}.

In recent years it has been shown that some nano-systems, for example the
semiconductor quantum-dots, coupled with a high-quality single-cavity-mode can
be a promising candidate for implementing the Hamiltonian of Dicke model and
the realistic parameter values are given by $\lambda\sim0.1-1.5$%
\cite{35,36,37} in the unit $\omega_{0}$. While $\omega_{0}$ is of order of
the underlying bulk band-gap with the value-range from 1.5 $eV$ in GaAs-like
semiconductor down to 0.1 $eV$ in narrow-gap semiconductor. The value of
$\lambda$ is also variable in a wide range from weak to strong coupling
regimes based on current nanotechnology. The GP can be observed directly by
measuring the mean photon number out of the cavity.

In conclusion, we have investigated the critical property of the non-adiabatic
GP derived for the first time in terms of time-dependent unitary
transformation in single-mode super-radiant Dicke model, which displays the
first-order QPT. The GP is shown to be proportional to the mean photon number,
and therefore is calculated analytically for the ground-state of the Dicke
model in the thermodynamical limit. We also provide the scaling behavior of GP
as a probe to test the QPT.

One of authors (G.C.) thanks Dr. S. L. Zhu and Dr. C. Emary for helpful
discussions and valuable suggestions. This work was supported by the Natural
Science Foundation of China under Grant No.10475053 and by the Natural Science
Foundation of Zhejiang Province under Grant No.Y605037.

\end{document}